\documentclass[11pt, conference, onecolumn]{IEEEtran}
\usepackage{amssymb}
\usepackage{subfigure}
\usepackage{amsmath}
\usepackage{amsfonts}
\usepackage{mathrsfs}
\usepackage{engord}
\usepackage[dvips]{graphicx}
\usepackage{threeparttable}
\usepackage{booktabs}
\usepackage{epsfig}
\usepackage{color}
\usepackage{graphicx}
\usepackage{multirow}
\usepackage{cite}
\usepackage{epstopdf}
\usepackage{framed}
\usepackage{url}
\usepackage{amsmath,bm}
\begin{document}

\newtheorem{thm}{Theorem}
\newtheorem{prop}{Proposition}
\newtheorem{lem}{Lemma}
\newtheorem{defn}{Definition}
\newtheorem{ex}{Example}
\newtheorem{cor}{Corollary}
\newtheorem{prn}{Principle}
\newtheorem{case}{Case}
%
\title{Chernoff Information of Bottleneck Gaussian Trees} 
\author{\IEEEauthorblockN{Binglin Li, Shuangqing Wei, Yue Wang, Jian Yuan}}
\maketitle
\let\thefootnote\relax\footnotetext{B. Li, Y. Wang and J. Yuan are with Department of Electronic Engineering, Tsinghua University,
Beijing, P. R. China, 100084. (E-mail: {libl13}@mails.tshinghua.edu.cn,  {wangyue, jyuan}@mail.tsinghua.edu.cn). S. Wei is
with the school of Electrical Engineering and Computer Science, Louisiana State University, Baton Rouge, LA 70803,
USA (Email: swei@lsu.edu).}

\begin{abstract}

In this paper, our objective is to find out the determining factors
of Chernoff information in distinguishing a set of Gaussian trees. In
this set, each tree can be attained via an edge removal and grafting
operation from another tree. This is equivalent to asking for the Chernoff
information between the most-likely confused, i.e. ``bottleneck", Gaussian
trees, as shown to be the case in ML estimated Gaussian tree graphs
lately.  We prove that the Chernoff information between two Gaussian
trees related through an edge removal and a grafting operation is the
same as that between two three-node Gaussian trees, whose topologies
and edge weights are subject to the underlying graph operation. In
addition, such Chernoff information is shown to be determined only
by the maximum generalized eigenvalue of the two Gaussian covariance
matrices.  The Chernoff information of scalar Gaussian variables as a
result of linear transformation (LT) of the original Gaussian vectors
is also uniquely determined by the same maximum generalized eigenvalue.
What is even more interesting is that after incorporating the cost of
measurements into a normalized Chernoff information,  Gaussian variables
from LT have larger normalized Chernoff information than the one based
on the original Gaussian vectors, as shown in our proved bounds.
\end{abstract}


\section{Introduction}

Gaussian graphical models have found great successes in
characterizing conditional independence of continuous
random variables in diverse applications including social
networks\cite{vega2007complex}, 
biology
\cite{ahmed2008time}, 
and economics\cite{dobra2010modeling}, to name a few. Among Gaussian
graphical models, Gaussian trees in particular have attracted
much attention due to their sparse structures, as well as existing
computationally efficient algorithms in learning the underling topologies
\cite{Chow.Liu.1968}.  The statistical inference problems related to
Gaussian graphical models are often focused on two primary aspects,
namely,  parameter estimation and performance analysis. The parameter
estimation is concerned of graph model selection and an estimation of the
associated covariance matrix. The focus of this paper is on the analysis
aspect, and more specifically, we want to develop some fundamental bounds
on Chernoff-Information (CI) based error exponents  in learning Gaussian
tree graphs. 


Chernoff information between two probability distributions offers
us an exact error exponent for the average error probability
in discerning the two distributions based on a sequence
of data drawn independently from one of the two distributions
\cite{cover2012elements}, and the minimum pair-wise Chernoff information
is the error exponent characterizing the performance of an $M$-ary
hypothesis testing problem \cite{westover2008asymptotic}. Recently in
\cite{santhanam2012information}, graph model selection problem has been
formulated as an information theoretical problem where each candidate
graph model is deemed as a message, and the sufficient conditions have
been found to decode and thus learn correctly the actual message (i.e. the
right graphical model). Such conditions were found by bounding pair-wise
error probabilities with some symmetrized distances between two candidate
graph models, which are weaker than Chernoff information, though.

Large deviation analysis has been conducted in
\cite{Tan.TSP.Ana.2010,Tan.TSP.Sang.2010,tan2011large} where the error
exponents for learning either discrete Markov or Gaussian trees
have been found. However, the error events in learning tree graph
models in \cite{Tan.TSP.Ana.2010,tan2011large} are conditional error
event, and a symmetrized Kullback--Leibler (KL) distance, namely,
{\em J--divergence}, instead of the tighter Chernoff distance, was
adopted in \cite{Tan.TSP.Sang.2010} to analyze the corresponding
error exponent. Such conditional error event was also considered in
\cite{jog2015model} where a tight lower-bound on KL distance between a
true Gaussian graph model and an incorrect one was found. In addition,
it was shown that such lower-bound is attained when the two graphs differ
by at least one edge, and the joint distribution of the candidate graph is
a projection of the true one onto it under the missing edge constraint.
It should be noted that it has been shown in  \cite{Tan.TSP.Ana.2010,
tan2011large} that the most likely error in ML estimation of a Markov
tree is another tree which differs from the true tree by a missing edge.

\begin{figure}
        \centering
        \includegraphics[width=7cm]{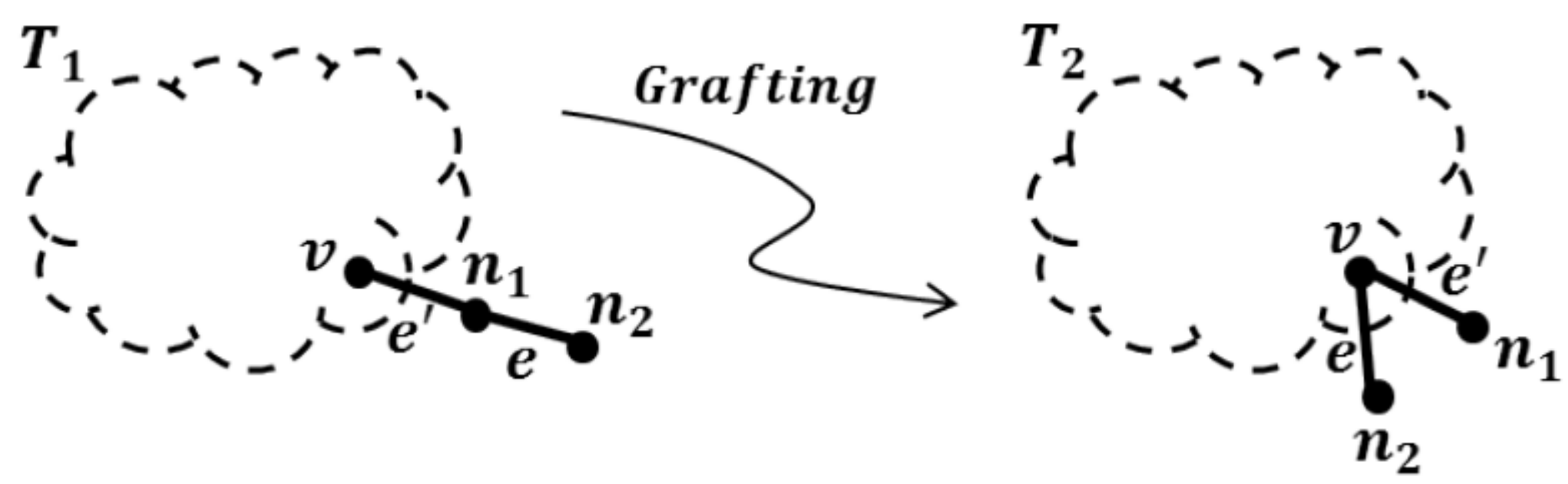}\\
        \caption{$T_2$ is obtained from $T_1$ by grafting operation}\label{grafting}
    \end{figure}

In this paper, we are interested in finding out the determining factors of
Chernoff information in distinguishing a set of Gaussian trees with the
same amount of randomness, i.e. sharing the same determinant of their
covariance matrices and normalized variances. In this set, any tree
in can be attained via some topological operation by grafting one edge
from another tree, as shown in Figure~\ref{grafting}.  Our formulation
is for the purpose of identifying the contributing factors to Chernoff
information in distinguishing Gaussian trees with minimum topological
difference, which can be deemed as a worst case scenario or equivalently
dominant error event as put in \cite{tan2011large}, with one edge missing
from one tree to get the other. The assumptions made on sharing the
same entropy and normalized variances are for the ease of analysis and
providing insights, as shown later in our results.

To reduce measurement cost, we could linearly transform an $N_I$
dimensional Gaussian vector ${\bf X}$ to a $N_O<N_I$ dimensional vector
${\bf Y} = {\bf A} {\bf X}$, through a $N_O$ by $N_I$ matrix ${\bf A}$.
An immediate question we address in this paper is the selection of ${\bf
A}$ through which we want to maximize the Chernoff information between
two two Gaussian distributions of ${\bf Y}$ corresponding to two original
Gaussian trees. In particular, we solved the problem for a simple, but
non-trivial case with $N_O=1$, i.e. the selection of a one by $N_I$
vector to maximize the Chernoff information between the two resulting
Gaussian scalars.

Our major and novel results can be summarized as follows. We first prove
a sequence of results on how to reduce the complexity of computing
Chernoff information between two Gaussian trees sharing some local
parameters. Based on these results, we further prove that the Chernoff
information between two Gaussian trees related through an edge removal
and a grafting operation is the same as that between two three-node
Gaussian trees, whose topologies and edge weights are subject to the
underlying graph operation. In addition, such Chernoff information is
determined only by the maximum generalized eigenvalue of the two Gaussian
covariance matrices.  The aforementioned transformation is further shown
to be applicable when we consider the Chernoff information between two
Gaussian scalar variables resulted from a linear transformation (LT)
of the original Gaussian vectors. 
What is even more interesting is that after incorporating the cost of
measurements into a normalized Chernoff information,  Gaussian variables
from LT have larger normalized Chernoff information than the one based
on the original Gaussian vectors, as shown in our proved bounds.


The paper is organized as follows. Section \ref{system model} presents
the system model. Some propositions used to simplify big trees are
presented in Section \ref{local}. Section \ref{3node} is about the study
of two simplified $3$-node trees and comparison of  how observation
cost affects Chernoff information.  And in Section \ref{conclusion}
we conclude the paper.

\section{System Model} \label{system model}
    Gaussian tree models capture the conditional independence relationships of multiple Gaussian variables using tree topologies.
    Here, we normalize the variance of all the values to be $1$ and all the mean values to be $0$. An $N$-node tree can be
represented as $\mathbf{G} = (V, E, W)$. Here $V=\{1,\dots,N\}$ is the vertex set of the tree. $E=\big\{e_{ij}|(i,j)\subset V\times
V\big\}$ is the edge set that satisfies $|E|=N-1$ and contains no cycles.
    $W=\{w_{ij}\in [-1,1]|e_{ij}\in E\}$ is the set of edge weights.
    On these conditions, a vector of Gaussian variables $\mathbf{x}=[x_1,x_2,\dots,x_N]^T\sim N(\mathbf{0},\Sigma)$ is said to be a Gaussian distribution on the tree $\mathbf{G} = (V, E, W)$ if
    \begin{align}
    \sigma_{ij}=
    \begin{cases}
    1&\quad i=j\\
    w_{ij}&\quad e_{ij}\in E\\
    w_{im}w_{mn}\dots w_{pj}&\quad e_{ij}\notin E
    \end{cases}
    \end{align}
     where $\sigma_{ij}$ is the $(i,j)$ term of $\Sigma$ and $e_{im}e_{mn}\dots e_{pj}$ is the unique path from node $i$ to node $j$\cite{moharrer2015classifying}.

Consider a set of Gaussian trees, namely, $\mathbf{G}_k=(V, E_k,
W_k)~k=1,2,\dots,m$, with their prior probabilities given by $\pi_1,\pi_2,
\dots, \pi_m$. They share the same entropy, and thus the same determinant
of their covariance matrices $\Sigma_k=[\sigma_{ij}^{(k)}]$, whose sets
of edge weights are denoted by $W_k=\{w_{ij}^{(k)}\}$.  We want to run
an $m$-ary hypothesis testing to find out from which Gaussian tree
the data sequence $\mathbf{X}=[\mathbf{x}_1,\dots,\mathbf{x}_T]$
($\mathbf{x}_l=[x_{1,l},\dots,x_{n,l}]'$) has been
drawn. We define the average error
probability of the hypothesis testing to be $P_e$, and let $E_e = \lim_{ T\rightarrow\infty}\frac{-\ln P_e}{T}$ be
the resulting error exponent\cite{cover2012elements},  which
depends on the smallest Chernoff information between the
trees \cite{westover2008asymptotic}, namely,
    \begin{align}
    E_e=\min_{1\leq i\neq j\leq m} CI(\Sigma_i||\Sigma_j)
    \end{align}
    where $CI(\Sigma_i||\Sigma_j)$ is the Chernoff information between the $i^{th}$ and $j^{th}$ trees.

    For two N-dim Gaussian joint distributions $\mathbf{x}_1\sim N(0,\Sigma_1)$ and $\mathbf{x}_2\sim N(0,\Sigma_2)$, the KL
distance from ${\bf G_1}$ to ${\bf G_2}$ is
    \begin{align}
    D(\Sigma_1||\Sigma_2)=&
    \frac{1}{2}\ln\frac{|\Sigma_2|}{|\Sigma_1|}+\frac{1}{2}tr(\Sigma_2^{-1}\Sigma_1)-\frac{N}{2}\label{D}
    \end{align}
    where $tr(\mathbf{A})=\sum a_{ii}$ is the trace of the matrix. We define a new distribution $N(0,\Sigma_\lambda)$ in the exponential family of the $N(0,\Sigma_1)$ and $N(0,\Sigma_2)$, namely
    \begin{align} \label{lambda.matrix}
    \Sigma_\lambda^{-1}=\Sigma_1^{-1}\lambda+\Sigma_2^{-1}(1-\lambda),
    \end{align}
\noindent and the Chernoff information is as given by
    \begin{align}
    CI(\Sigma_1||\Sigma_2)=D(\Sigma_{\lambda^*}||\Sigma_2)=D(\Sigma_{\lambda^*}||\Sigma_1),
    \end{align}
\noindent    where $\lambda^*$ is the point which satisfies the latter equation\cite{cover2012elements}.
    As expected,  the computational complexity of Chernoff information
    greatly depends on the two specific trees involved.

It has been shown recently in \cite{jog2015model} that the KL distance
between a true Gaussian graph model and an incorrect one is lower-bounded
by a conditional mutual information, and the lower-bound becomes tight
when the learned tree only differs from the true one by removal of one
leaf node and grafting to another vertex in the original tree.  As we
already know that the overall Chernoff information in an $m$-ary testing
is bottle-necked by the minimum pair-wise difference, thus for the rest
of this paper, we only consider pairs of Gaussian trees one of which can
be obtained from the other through such grafting operations in order to
reveal what really determines the Chernoff information related to such
dominant error events.

In addition to the full observation case, we will also study a
    linearly transformed (LT) observation case. For two $N$-node trees:
    $\mathbf{G}_1,\mathbf{G}_2$, in the full
    observation case, we can have access to all $N$ variables and only need
    to calculate $CI(\Sigma_1||\Sigma_2)$. But in the LT observation
    case, we can only observe a $N_O$-dim vector each time, namely,
    $\mathbf{y}=\mathbf{A}\mathbf{x}$, where $\mathbf{A}$ is a $N_O\times
    N$ matrix and $\mathbf{x}\in R^N,\mathbf{y}\in R^{N_O}$. The new
    variables follow joint distributions $N(\mathbf{0},\tilde{\Sigma}_1)$
    and  $N(\mathbf{0},\tilde{\Sigma}_2)$. For fixed $N_O$, we want to
    find the optimal $\mathbf{A}^*$ and its Chernoff information result
    $CI(\tilde{\Sigma}_1^*||\tilde{\Sigma}_2^*)$, s.t.
    \begin{align}
    \mathbf{A}^*=\arg \max_{\mathbf{A}} CI(\tilde{\Sigma}_1||\tilde{\Sigma}_2)\label{2A}
    \end{align}

\noindent    For $m$-ary Hypothesis testing case, the optimal $\mathbf{A}^*$ becomes
    \begin{align}
    \mathbf{A}^*=\arg \max_{\mathbf{A}} \min_{1\leq i\neq j\leq m} CI(\tilde{\Sigma}_i||\tilde{\Sigma}_j)\label{mA}
    \end{align}

    To gain insights, we only consider and compare two cases, the full
    observation one against  the case of $N_O=1$ for the LT
in terms of their respective Chernoff information with and without a
normalization factor to count the differences reflected by measurement
dimensions.   Next section, we will provide some interesting results
to shed
    light on the determining factors of Chernoff information between
    two trees with minimum structure difference.

\section{Propositions to simplify the calculation of Chernoff information} \label{local}

    A normalized covariance matrix of a Gaussian tree has a very simple inverse matrix and determinant, which are necessary for the calculation of Chernoff information.

    \begin{prop}\label{thm1}
    For a normalized covariance matrix of Gaussian trees  $G=(E,V,W)$, define $\Sigma^{-1}=[u_{ij}]$. So $|\Sigma|=\prod_{e_{ij}\in E}(1-w_{ij}^2)$ and the elements of $\Sigma^{-1}$ follow the following expressions:
    \begin{align}
    u_{ij}=
    \begin{cases}
    \frac{-w_{ij}}{1-w_{ij}^2}&\quad i\neq j~\text{and}~e_{ij}\in E\\
    0&\quad i\neq j~\text{and}~e_{ij}\notin E\\
    1+\sum_{p:e_{ip}\in E}\frac{w_{ip}^2}{1-w_{ip}^2}& \quad i=j.
    \end{cases}
    \end{align}
    \end{prop}

    This proposition can be easily proved with the equation of block matrix and mathematical induction.

\subsection{Full observation case}

    For the full observation case, we can observe all the $N$ nodes
    each time. Then we can learn the resolution potential of the trees
    without the effect of observation mapping. Here we learn how small
    local differences influence the Chernoff information.  The grafting
    operation contains cutting operation and attaching operation.
    We want to study the cutting operation first and see how  Chernoff
    information changes when we cut the same vertex from two trees.
    For convenience, we define a new symbol $A=B/\{b\}$, indicating that
 the    set $A$ has all the elements in set $B$ except element $b$.

    \begin{prop}\label{combine-graph}
    For two $N$-node Gaussian graphical models $\mathbf{G}_1=(V, E_1, W_1),\mathbf{G}_2=(V, E_2, W_2)$, their Chernoff information is written as $CI(\mathbf{G}_1||\mathbf{G}_2)$. Then we draw two new Gaussian graphical models $\mathbf{G}_1'=(V/\{i\}, E_1', W_1'),\mathbf{G}_2'=(V/\{i\}, E_2', W_2')$ whose joint distributions are the same with the  joint distribution of $V/\{i\}$ nodes in $\mathbf{G}_1,\mathbf{G}_2$. Their Chernoff information is written as $CI(\mathbf{G}_1'||\mathbf{G}_2')$, with $CI(\mathbf{G}_1'||\mathbf{G}_2')\leq CI(\mathbf{G}_1||\mathbf{G}_2)$.
    \end{prop}

    To prove this proposition, we have to prove several other results first.

    \begin{prop}\label{combine-states}
    Assuming that there are two $n$ states discrete PMF
    $\mathbf{P}=[p_1,p_2,\dots,p_{n-2},p_{n-1},p_n]$ and $\mathbf{Q}=[q_1,q_2,\dots,q_{n-2},q_{n-1},q_n]$. We combine the ${(n-1)}^{th} , {n}^{th}$
    states and get two new PMF $\mathbf{P}'=[p_1,p_2,\dots,p_{n-2},p_{n-1}+p_n]$ and  $\mathbf{Q}'=[q_1,q_2,\dots,q_{n-2},q_{n-1}+q_n]$. Then $CI(\mathbf{P}||\mathbf{Q})\geq CI(\mathbf{P}'||\mathbf{Q}')$ and the equation holds if and only if $\frac{p_{n-1}}{q_{n-1}}=\frac{p_{n}}{q_{n}}$ or $q_{n-1}=q_n=0$.
    \end{prop}

    \begin{prop} \label{combine-continous}
    Assume that there are $N$ continuous values $\mathbf{X}=[X_1,\dots,X_N]$ and two distributions $f_1(\mathbf{X})$ and  $f_2(\mathbf{X})$, their joint distributions on nodes $\mathbf{X}/\{i\}$ are $f_1'(\mathbf{X}/\{i\})$ and $f_2'(\mathbf{X}/\{i\})$. So the Chernoff information between them follows this property:
    \begin{align}
    CI(f_1(\mathbf{X})||
    f_2(\mathbf{X}))\geq
    CI(f_1'(\mathbf{X}/\{i\})||
    f_2'(\mathbf{X}/\{i\}))
    \end{align}
    and it becomes equality if and only if  the conditional distribution
    follows $f_1(X_i|\mathbf{X}/\{i\})=f_2(X_i|\mathbf{X}/\{i\})$.
    \end{prop}

    Proposition \ref{combine-states} tells us that combining  states will not increase  the Chernoff information between distributions. We use the Holder inequality and the Chernoff information computation via exponential family to prove it.
    To prove proposition \ref{combine-continous}, we should
    discretize these continuous variables into discrete ones. Then
    we can use proposition \ref{combine-states} to prove proposition
    \ref{combine-continous} directly. Proposition \ref{combine-graph}
    is the graph version of proposition \ref{combine-continous}.

    Propositions \ref{combine-continous} is not only an inequality like proposition \ref{combine-graph}. It also tells us cutting what kinds of nodes doesn't change the Chernoff information. We can ignore these nodes without performance loss in terms of Chernoff information.

    \begin{prop}\label{thm2}
    For two $N$-node Gaussian tree models $\mathbf{G}_1=(V, E_1, W_1)$ and $\mathbf{G}_2=(V, E_2, W_2)$, their Chernoff information is written as $CI(\mathbf{G}_1||\mathbf{G}_2)$. If they have the same leaf node $i$ with an edge connecting to the same node $j$ with the same weight $w_{ij}$, then we can delete this node and edge and get two new trees $\mathbf{G}_1'=(V/\{i\}, E_1/e_{ij}, W_1/w_{ij}),\mathbf{G}_2'=(V/\{i\}, E_2/e_{ij}, W_2/w_{ij})$, with $CI(\mathbf{G}_1'||\mathbf{G}_2')= CI(\mathbf{G}_1||\mathbf{G}_2)$.
    \end{prop}

    \begin{prop}\label{thm3}
    For two $N$-node Gaussian tree models $\mathbf{G}_1=(V, E_1, W_1)$ and $\mathbf{G}_2=(V, E_2, W_2)$, their Chernoff information is written as $CI(\mathbf{G}_1||\mathbf{G}_2)$. If they have the same internal node $i$ with two edges connecting to the same nodes $p,q$ with the same weights $w_{ip},w_{iq}$, then we can delete this node and edges, add new edge $e_{pq}$ with weight $w_{ip}w_{iq}$ instead, and get two new trees $\mathbf{G}_1'=(V/\{i\}, E_1', W_1'),\mathbf{G}_2'=(V/\{i\}, E_2', W_2')$, with $CI(\mathbf{G}_1'||\mathbf{G}_2')= CI(\mathbf{G}_1||\mathbf{G}_2)$.
    \end{prop}

\noindent    These two propositions are direct extensions of proposition
    \ref{combine-continous}.  If we have a
    $1$-degree or $2$-degree node with identical local relationship and
    correlation parameters, we can remove the leaf or combine its two
    edges without changing the Chernoff information.
     We consider  nodes with degree $1$ and $2$, because only in these
     situations are new graphs still trees.

    Note that we only consider the Chernoff information of two Gaussian
    trees differing through a single edge grafting operation.  In other
    words, the two Gaussian trees correspond to two $N$ by $N$ covariance
    matrices $\Sigma_1$ and $\Sigma_2$,  sharing the same determinant, and
    with normalized variances. In particular, $G_2$ is attained from $G_1$
    by cutting one edge connecting node $i$ and $j$ from any node $i$
    and then connecting node $j$ to another node $k$, as shown in Figure~\ref{grafting},  without changing the
    weights of all edges in order to maintain the same determinant, which is determined by the product of $(1-|\omega_{ij}|^2)$, for
all edges of $(i,j) \in E$.  We can use proposition \ref{thm2} and \ref{thm3}
    repeatedly to reduce the original $N$-node trees $\mathbf{G}_1$
    and $\mathbf{G}_2$ into two special $3$-node trees $\mathbf{G}_1'$
    and $\mathbf{G}_2'$ shown as Figure~\ref{3node-p}. In the new trees,
    $w_2 = w_{ij}$ (the edge weight of $e_{ij}$), and $w_1$ is the weight
    of the path connecting node $i$ with node $k$ in the graph $G_1$,
    i.e. $w_1 = \prod_{ e \in \text{path} ~ik}w_{e}$.

\subsection{$1$-dim LT observation in $2$-ary Hypothesis testing}

    For the full observation case, we know that nodes with the same
    local subgraph relationship do not making contribution to
Chernoff information. It is thus anticipated that this type of nodes does
not provide any gain when we can only observe a $1$-dim LT mapping output.
    \begin{prop}\label{thm4}
    For two $N$-node Gaussian tree models $\mathbf{G}_1=(V, E_1,
    W_1)$ and $\mathbf{G}_2=(V, E_2, W_2)$, we can't get all
    the data of the values but only one linearly combining value
    $\mathbf{Y}=\boldsymbol{\alpha}\mathbf{X}$ where $\boldsymbol{\alpha}$
    is an $1\times N$ vector, and  $\boldsymbol{\alpha}^*= \arg \max
    CI(\mathbf{Y}^{(1)}||\mathbf{Y}^{(2)})$  is the optimal observation
    mapping. If node $i$ has identical local relationship and correlation
    parameters as shown in proposition \ref{thm2} and \ref{thm3},
    there exists an optimal $\boldsymbol{\alpha}^*$ whose $i^{th}$
    value $\alpha^*_i$ equals to $0$.  \end{prop}

\noindent It tells us that we needn't observe the nodes if the
local subgraph around it is completely the same. We only need to see
a linear combination of  other nodes to get enough information for
distinguishing. Using this proposition, we can leave out the same
subgraphs of two trees and focus on the different parts when we are
looking for the optimal observation.

\subsection{The comparison between two observation models}
    Intuitively, the Chernoff information in full observation is always
    larger than that of LT observation as a consequence of data processing
    inequality \cite{cover2012elements}.

    \begin{prop}\label{>1}
    For two $N$ states distributions $\mathbf{X}_{1}$ and $\mathbf{X}_{2}$ of two graphs, we can use two different observation matrices and get different output $\mathbf{Y}_p=\mathbf{P}_{p\times N}\mathbf{X}$ and $\mathbf{Y}_q=\mathbf{Q}_{q\times N}\mathbf{X}$ where $q<p\leq N$. $\mathbf{P}^*$ and $\mathbf{Q}^*$ are the optimal matrix shown as (\ref{mA}) under the observation constraint. So $CI(\mathbf{Y}_p^{(1)*}||\mathbf{Y}_p^{(2)*})\geq CI(\mathbf{Y}_q^{(1)*}||\mathbf{Y}_q^{(2)*})$.
    \end{prop}

    The more information we have access to, the larger optimal Chernoff
    information will be. Due to the LT operation, we have lower-dim
    observation data. The dimension-reduced result will decrease
    the Chernoff information. However, we are interested in a more
    fair comparison  between Chernoff information after counting their
    observation dimensions. Particularly and surprisingly, LT observation
    yields larger normalized the Chernoff information than the case with
    full observation, as shown next in Section~\ref{compare}.

\subsection{An important parameter $\lambda_{max}$}\label{lambda}

We next introduce a critical parameter $\lambda_{max}$, which will play
a key role in determining the Chernoff information of two Gaussian trees
between which there is only minor difference in topology and parameters
due to the edge-wise grafting operation, entailing the underlying error
event the dominant one in distinguishing between one of two trees with
those ``adjacent" ones.

\begin{prop}
    For the two trees $\mathbf{G}_1'$ and $\mathbf{G}_2'$ in Figure~\ref{3node-p}, the generalized eigenvalue of their correlation matrix $\Sigma_1$ and $\Sigma_2$, i.e.
    the eigen-values of the resulting matrix $S= \Sigma_1 \Sigma_2^{-1}$ is $\{1, \lambda_{max}, 1/\lambda_{max} \}$ with $\lambda_{max}$ determined by
    \begin{gather}
    \lambda_{max} = \frac{\sqrt{\beta} + w_2}{\sqrt{\beta} - w_2}
    \end{gather}
    \noindent where $\beta = w_2^2 + 2 \frac{1- w_2^2}{1-w_1}$, and
    $w_2$ is the weight of cut edge,  and $w_1$ is the weight of the path connecting the neighboring vertices of the moved node
before and after the grafting operation.
    \end{prop}

\noindent Its proof hinges upon the symmetric property of the Chernoff
information, as well as the shared determinant of the matrices
$\Sigma_1$ and $\Sigma_2$.


\section{Chernoff information of $3$-node Gaussian trees}\label{3node}

\begin{figure}
      \centering
      \includegraphics[width=4cm]{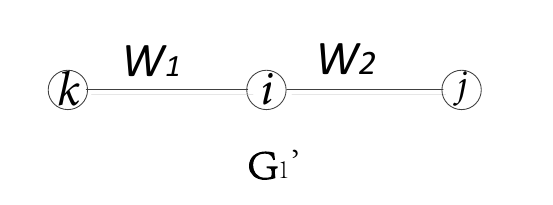}
        \includegraphics[width=3cm]{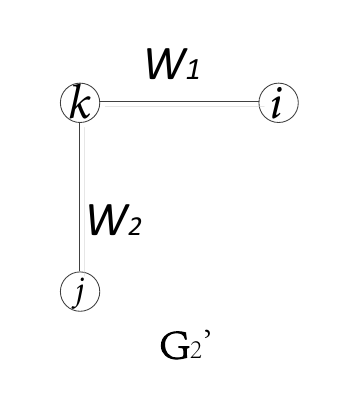} \\
      \caption{Special $3$-node case}\label{3node-p}
    \end{figure}

    Propositions \ref{thm2}, \ref{thm3} and \ref{thm4} show that we can
    transform two large and similar Gaussian trees into two  $3$-node
    trees when calculating their Chernoff information, as shown in
    Figure~\ref{3node-p}. Thus, in this section, we shift our attention
    to the Chernoff information of two such $3$-node trees. We want
    to calculate their Chernoff information in full observation case
    and $1$-dim LT observation case. And then we will compare these
    two Chernoff information with a  fair metric in terms of Chernoff
    information per measurement dimension.

\subsection{$1$-dim LT observation in $2$-ary Hypothesis testing}\label{calculate1}

    For two Gaussian tree models shown in figure \ref{3node-p}, we can only observe an $1$-dim linear output $\mathbf{Y}=\boldsymbol{\alpha}\mathbf{X}$ each time, where $\boldsymbol{\alpha}$ is an $1\times3$ observation vector and $\mathbf{Y}$ is an $1$-dim output distribution. We want to find the optimal observation vector $\boldsymbol{\alpha}^*$ to maximum the Chernoff information between the output of two trees. So
    the optimal LT mapping vectors are $\boldsymbol{\alpha}_1=k[s_1,s_2,1],\boldsymbol{\alpha}_2=k[s_2,s_1,1]$, and the maximum
Chernoff information $CI_1(\Sigma_1||\Sigma_2)$ is $g(\lambda_{max})$, where $g(x)$ is shown in (\ref{g}),  $k$ is an arbitrary
non-zero number and $s_1=-\frac{1}{2}\left(w_2+\sqrt{\beta}\right)$, $
    s_2=
    -\frac{1}{2}\left(w_2-\sqrt{\beta}\right)$.
    \begin{align}
    g(x)=&\frac{1}{2}\left\{\ln\frac{x-1}{e\ln x}+\frac{\ln x}{x-1}\right\}\label{g}
    \end{align}
    $\beta$ and $\lambda_{max}$ is defined in section \ref{lambda}.
    The proof exploits the concavity property  of the transformed
    objective function which is subject to the ratio of the variances
    of two scalar Gaussian variables after the LT operation. Its  maximum
    value is further expressed using the maximum generalized eigenvalue
    of $\Sigma_1$ and $\Sigma_2$.

\subsection{Full observation case}\label{calculate2}
    In this case, the Chernoff information between $\mathbf{G}_1'$ and $\mathbf{G}_2'$ shown in figure \ref{3node-p} is
    \begin{align}
    CI_2(\Sigma_1||\Sigma_2)=\ln(\frac{\lambda_{max} +1}{2 \sqrt{\lambda_{max}}})\label{CI2}
    \end{align}

\noindent The main idea of the proof is to take advantage of the
topological relationship between the two trees in Figure~\ref{3node-p},
and then seek the optimal covariance matrix in the form of
Eq.~\ref{lambda.matrix}.

\subsection{Comparing the two Chernoff information}\label{compare}
\begin{figure}
      \centering
      \includegraphics[width=7cm]{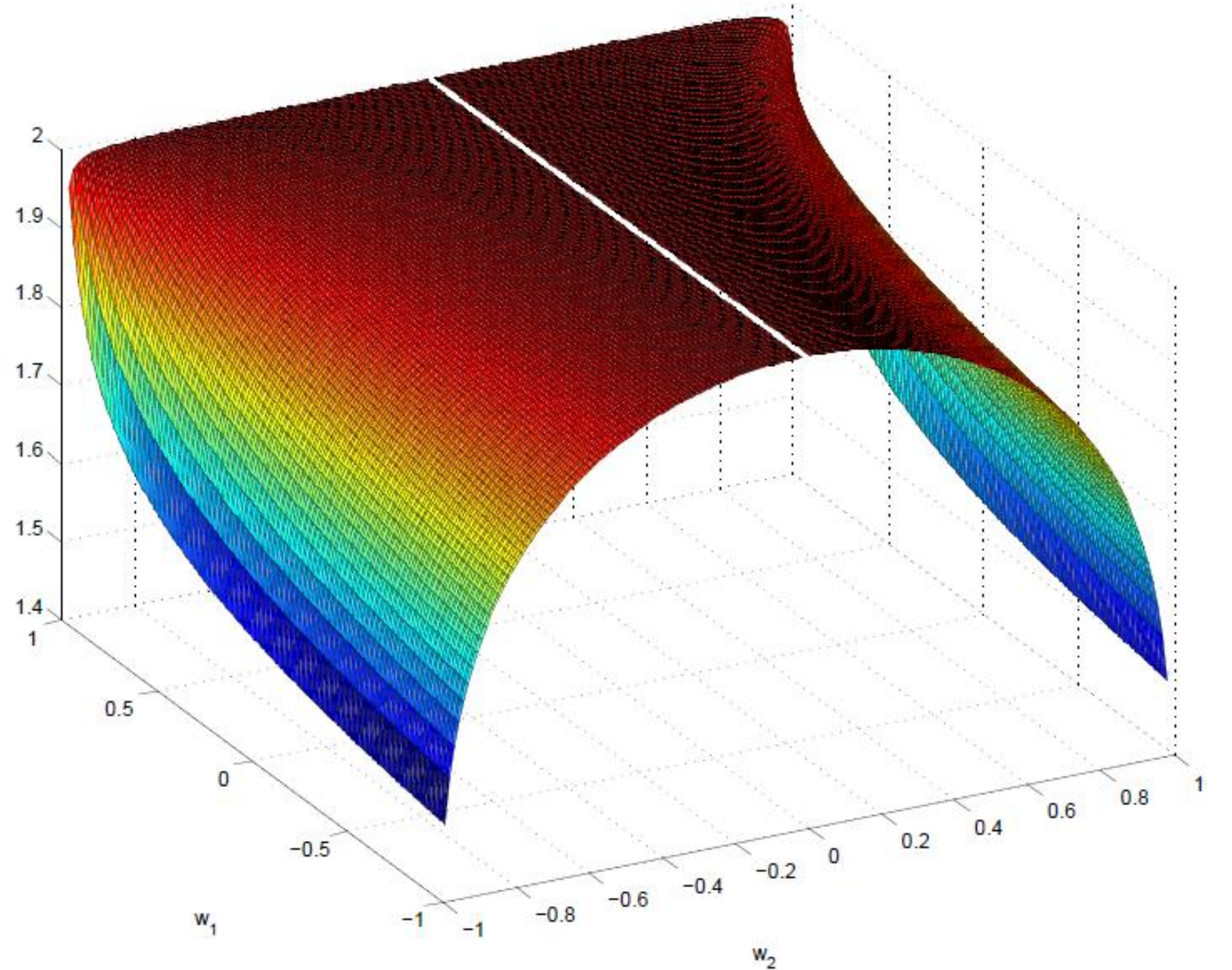}\\
      \caption{Figure of $\frac{CI_2}{CI_1}$}\label{plot}
    \end{figure}

    In the case of full observation and $1$-dim LT observation,
    Figure~\ref{plot} shows the ratio of their Chernoff information
    expressed with parameters of $(w_1,w_2)$.

\begin{prop}\label{<2}
    For two Gaussian trees $\mathbf{G}_1$ and $\mathbf{G}_2$, where
    $\mathbf{G}_1$ can be obtained from $\mathbf{G}_2$ by a single edge grafting operation,
    the ratio of their Chernoff information under full
    observation and $1$-dim LT observation satisfies:
    $1\leq\frac{CI_2}{CI_1}\leq2$.
\end{prop}

    The ratio must not be greater than $1$ because of proposition
    \ref{>1}. And from equation (\ref{g}), (\ref{CI2}) and
    $\lambda_{max}\geq1$, we can prove that $\frac{CI_2}{CI_1}\leq2$.

    Given two trees $\mathbf{G}_1$ and $\mathbf{G}_2$, the error
    probability $P_{e1} \leq e^{-T CI_1}$ in $1$-dim LT observation
    testing and $P_{e2} \leq e^{-T CI_2}$ in full observation testing,
    where $T$ is the number of slots expended for testing purpose.
    However in full observation case, we have three effective measurement
    values in each slot, and thus we actually have $\hat{T}_2=3 T$ number
    of measurements collected in total. As a contrast, the LT approach
    is based on $\hat{T}_1= T$ measurements only.  To fairly compare
    the two approaches, we propose a normalized Chernoff information
$\hat{CI}_j = \lim_{\hat{T}_j \rightarrow \infty}
\frac{-\log(P_{ej})}{\hat{T}_j}$, where $\hat{T}_j$ denotes the total
number of real valued measurements collected.

As a result, the normalized Chernoff information for the two cases are:
$\hat{CI}_2 = \frac{CI_2}{3}$, and $\hat{CI}_1 = CI_1$, which implies that
the ratio of such normalized Chernoff information satisfies: $1/3 \leq
\frac{\hat{CI}_2}{\hat{CI}_1} \leq 2/3$.  We can see that the LT actually
has a larger normalized Chernoff information than that with full access to
the original data, which suggests its efficiency after we have taken the
dimension of the sample size into consideration. This is quite surprising,
considering that linear mapping always reduces Chernoff information
without putting normalization into the picture, and it thus demonstrates
a favor towards the LT option  after we count the measurement cost into
the comparison of Chernoff information between the two approaches.

\section{Conclusion}\label{conclusion}

In this paper, we have shown how local changes in topology and
parameters affect the capacity of distinguishing two resulting Gaussian
trees measured by Chernoff information. In particular, the maximum
generalized eigenvalue $\lambda_{max}$ is shown to play a critical role in
determining the Chernoff information with or without linear transformation
mapping. Our proposed normalized Chernoff information is able to reflect
the discerning capability in terms of error exponents with a constraint
of the same amount of measurement cost. In one of our future works,
we will investigate how Chernoff information varies if a Gaussian tree
is attained from the other one via a sequence of local graph operations.

\bibliography{1201}
\bibliographystyle{IEEEtran}

\newpage
\appendices

 \section{Proof of Propositon~\ref{thm1}}

     We use the equation of block matrix and mathematical induction to prove them.

    1) For a 2-node tree, $\Sigma=\begin{bmatrix} 1&w\\w&1 \end{bmatrix}$, $|\Sigma|=1-w^2$ and $\Sigma^{-1}=\begin{bmatrix} 1+\frac{w^2}{1-w^2}&\frac{-w}{1-w^2}\\\frac{-w}{1-w^2}&1+\frac{w^2}{1-w^2} \end{bmatrix}$.

    2) For an arbitrarily $N$-node tree with the normalized $\Sigma$, assume  $\Sigma^{-1}=\begin{bmatrix} A_{(N-1)*(N-1)}&B_{(N-1)*1}\\B^T&a \end{bmatrix}$
     and \\$\Sigma=\prod_{e_{ij}\in E}{(1-w_{ij}^2)}$. At the same time, we can set the last column of $\Sigma$ as $\boldsymbol{\alpha}$, which satisfies $\Sigma^{-1}\boldsymbol{\alpha}=[0,0,\dots,0,1]^T$.

     A $(N+1)$-node tree can be treated as a $N$-node tree added with a new leaf. Without loss of generality, the $N$-node tree has node $1,2,\dots,N$ and the new edge is $e_{N,N+1}$ with parameter $w$.  So
     \begin{align}
     \Sigma'=\begin{bmatrix} \Sigma &w\boldsymbol{\alpha}\\w\boldsymbol{\alpha}^T&1 \end{bmatrix}
     \end{align}
     \begin{align}
     |\Sigma'|=&|\Sigma||1-w^2\boldsymbol{\alpha}^T\Sigma^{-1}\boldsymbol{\alpha}|
     \nonumber\\=&|\Sigma|(1-w^2\boldsymbol{\alpha}(N))
     \nonumber\\=&|\Sigma|(1-w^2)
     \nonumber\\=&\prod_{e_{ij}\in E'}{(1-w_{ij}^2)}
     \end{align}
     \begin{align}
     \Sigma'^{-1}=&
     \begin{bmatrix} \Sigma^{-1}+\frac{w^2}{1-w^2}\Sigma^{-1}\boldsymbol{\alpha}\boldsymbol{\alpha}^T \Sigma^{-1} &\frac{-w}{1-w^2}\Sigma^{-1}\boldsymbol{\alpha}\\
     \frac{-w}{1-w^2}\boldsymbol{\alpha}^T\Sigma^{-1}&\frac{1}{1-w_N^2} \end{bmatrix}
     \nonumber\\=&
     \begin{bmatrix}
     A&B&\boldsymbol{0}_{(N-1)*1}\\
     B^T&a+\frac{w^2}{1-w^2}&\frac{-w}{1-w^2}\\
     \boldsymbol{0}_{1*(N-1)}&\frac{-w}{1-w^2}&1+\frac{w^2}{1-w^2}
     \end{bmatrix}
     \end{align}

     So the $(N+1)$-node tree fits the supposition and the proposition is proved.

 \section{Proof of Proposition~\ref{combine-states}}

    We define the energy function of the two exponential family as below:

    \begin{align}
    F(\lambda)=\sum_{i=1}^{n-2}p_i^\lambda q_i^{1-\lambda}+
    p_{n-1}^\lambda q_{n-1}^{1-\lambda}+
    p_{n}^\lambda q_{n}^{1-\lambda}\\
    F'(\lambda)=\sum_{i=1}^{n-2}p_i^\lambda q_i^{1-\lambda}+
    {(p_{n-1}+p_n)}^\lambda{(q_{n-1}+q_n)}^{1-\lambda}
    \end{align}
    So the Chernoff information is a function of $F(\lambda)$:
    \begin{align}
    CI(\mathbf{P}||\mathbf{Q})=-\min_{0\leq\lambda\leq1}\ln{F(\lambda)}\\
    CI(\mathbf{P}'||\mathbf{Q}')=-\min_{0\leq\lambda\leq1}\ln{F'(\lambda)}
    \end{align}

    $p_{n-2}^\lambda q_{n-2}^{1-\lambda}+
    p_{n}^\lambda q_{n}^{1-\lambda}\leq {(p_{n-1}+p_n)}^\lambda{(q_{n-1}+q_n)}^{1-\lambda}$  when $0<\lambda<1$ because of the Holder Inequality, so $F(\lambda)\leq F'(\lambda)$ and the equation holds if and only if $\frac{p_{n-1}}{q_{n-1}}=\frac{p_{n}}{q_{n}}$ or $q_{n-1}=q_n=0$.

    a) When $\frac{p_{n-1}}{q_{n-1}}=\frac{p_{n}}{q_{n}}$ or $q_{n-1}=q_n=0$, $\ln {F(\lambda)} = \ln{F'(\lambda)}$ and therefore $CI(\mathbf{P}||\mathbf{Q})=CI(\mathbf{P}'||\mathbf{Q}')$.

    b) Otherwise we can get $\ln{F(\lambda)} < \ln{F'(\lambda)} \leq 0$ when $0<\lambda<1$ and $\ln{F(\lambda)} = \ln{F'(\lambda)}=0$ when $\lambda=0,1$. So $\min_{0\leq\lambda\leq1}\ln{F(\lambda)}< \min_{0\leq\lambda\leq1}\ln{F'(\lambda)}$ and $CI(\mathbf{P}||\mathbf{Q})>CI(\mathbf{P}'||\mathbf{Q}')$.

    In summary $CI(\mathbf{P}||\mathbf{Q})\geq CI(\mathbf{P}'||\mathbf{Q}')$ and the equation holds if and only if $\frac{p_{n-1}}{q_{n-1}}=\frac{p_{n}}{q_{n}}$ or $q_{n-1}=q_n=0$.

 \section{Proof of Proposition~\ref{combine-continous}}

    We divide the range of the values $X_1,X_2,\dots,X_N$ into $K_1,K_2,\dots,K_N$ pieces. So $X_i$ has $A=K_i$ states,  $\mathbf{X}/\{i\}$ has $B=\prod_{m=1,m\neq i}^N K_m$ states, and $\mathbf{X}$ has $C=AB$ states.

    $f_1(\mathbf{X})$ and  $f_2(\mathbf{X})$ can be treated as two $1$-dim discrete distributions $P(\mathbf{X})$ and $Q(\mathbf{X})$ with $C$ possible states when  $K_1,K_2,\dots,K_N$ are large enough. And $f_1'(\mathbf{X}/\{i\})$ and $f_2'(\mathbf{X}/\{i\})$ can be treated as two a-dim discrete distributions $P'(\mathbf{X}/\{i\})$ and $Q'(\mathbf{X}/\{i\})$ with $B$ possible states.

    Observe the states of $P(\mathbf{X})$ and $P'(\mathbf{X}/\{i\})$ and we can find their relationship. If we combine all the $A$ states with the same $X_i$, $P(\mathbf{X})$ with $C$ states will become $P'(\mathbf{X}/\{i\})$ with $B$ states. Using proposition \ref{combine-states} repeatedly, we will get
    $CI(\mathbf{P}||\mathbf{Q})\geq CI(\mathbf{P}'||\mathbf{Q}')$, and the equation holds  if and only if $P(X_i|\mathbf{X}/\{i\})=Q(X_i|\mathbf{X}/\{i\})$.

    When $K_1,K_2,\dots,K_N$ goes to infinite, $P(\mathbf{X})$ and $Q(\mathbf{X})$ go to $f_1(\mathbf{X})$ and  $f_2(\mathbf{X})$. $P'(\mathbf{X}/\{i\})$ and $Q'(\mathbf{X}/\{i\})$ go to $f_1'(\mathbf{X}/\{i\})$ and $f_2'(\mathbf{X}/\{i\})$. The inequality remains in the process of convergence. So
    \begin{align}
    CI(f_1(\mathbf{X})||
    f_2(\mathbf{X}))\geq
    CI(f_1'(\mathbf{X}/\{i\})||
    f_2'(\mathbf{X}/\{i\}))
    \end{align}
    and it becomes equality if and only if  the conditional distribution
    follows $f_1(X_i|\mathbf{X}/\{i\})=f_2(X_i|\mathbf{X}/\{i\})$.

\section{Proof of Proposition~\ref{thm4}}

    The relative entropy of two zero-mean Gaussian distribution $N(0,\sigma^2_1)$ and $N(0,\sigma^2_1)$ is
    \begin{align}
    D(\sigma_1^2||\sigma_2^2)=&\int_{-\infty}^\infty f_{\sigma_1^2}(x)\log\frac{f_{\sigma_1^2}(x)}{f_{\sigma_2^2}(x)}dx\nonumber\\
    =&\frac{1}{2}\log\frac{\sigma_2^2}{\sigma_1^2}+\frac{\sigma_1^2}{2\sigma_2^2}-\frac{1}{2}
    \end{align}
    The Chernoff information between them is
    \begin{align}
    CI(\sigma_1^2||\sigma_2^2)=
    D(\sigma'^2||\sigma_1^2)=D(\sigma'^2||\sigma_2^2)
    \end{align}
    where
    \begin{align}
    \sigma'^2=\log\frac{\sigma_2^2}{\sigma_1^2}
    \frac{\sigma_1^2\sigma_2^2}{\sigma_2^2-\sigma_1^2}
    \end{align}
     So
    the Chernoff information is a function of $\frac{\sigma_2^2}{\sigma_1^2}$, namely,
    \begin{align}
    CI(\sigma_1^2||\sigma_2^2)=g(\frac{\sigma_2^2}{\sigma_1^2})\label{1dimCI}\\
    g(x)=\frac{1}{2}\{\log\frac{x-1}{e\log x}+\frac{\log x}{x-1}\}
    \end{align}
    $g(x)$ is a increasing function when $x>1$, and satisfies $g(x)=g(\frac{1}{x})$.

    If we want to maximize the Chernoff information, we can maximize the proportion between the bigger variance and the smaller variance instead.

    Then we provide a simple proposition at first and we will use it repeatedly later. It is easy to prove.

    \begin{prop}\label{fact}
    If $B_i>0,a\leq\frac{A_i}{B_i}\leq b(i=1,\dots,k),\sum c_iB_i>0$, so
    $a\leq\frac{\sum c_iA_i}{\sum c_iB_i}\leq b$.
    \end{prop}

    We prove proposition \ref{thm4} in two different cases.

\subsection{Node $i$ is a leaf with the same neighbor}

    Without loss of generality, we make $i=N,w_{ij}=w$.

    The $N-1$ node trees without node $N$ have the normalized covariance matrix  $\Sigma_1$ and $\Sigma_2$. We set the $j$-st column of $\Sigma_1$ as $\boldsymbol{\beta}_1$, so as $\boldsymbol{\beta}_2$.

     The complete trees can be treated as the $N-1$ node trees added with a new leaf $N$. So
     \begin{align}
     \Sigma_1'=\begin{bmatrix} \Sigma_1 &w\boldsymbol{\beta}_1\\w\boldsymbol{\beta}_1^T&1 \end{bmatrix}\\
     \Sigma_2'=\begin{bmatrix} \Sigma_2 &w\boldsymbol{\beta}_2\\w\boldsymbol{\beta}_2^T&1 \end{bmatrix}
     \end{align}

     We can find $x\geq1$ and $\boldsymbol{\alpha}^*$ so that $\frac{1}{x}\leq\frac{\boldsymbol{\alpha}^T\Sigma_1\boldsymbol{\alpha}}
     {\boldsymbol{\alpha}^T\Sigma_2\boldsymbol{\alpha}}\leq x$ holds for arbitrary $\boldsymbol{\alpha}$ and $\frac{\boldsymbol{\alpha}^{*T}\Sigma_1\boldsymbol{\alpha}^*}
     {\boldsymbol{\alpha}^{*T}\Sigma_2\boldsymbol{\alpha}^*}=x \:\text{or}\: \frac{1}{x}$. As (\ref{1dimCI}) shown,   $\boldsymbol{\alpha}^*$ is the optimal observation of the $N-1$-node trees and the optimal Chernoff information is $g(x)$.

     We want to prove that $\frac{1}{x}\leq\frac{\boldsymbol{\gamma}^T\Sigma_1'\boldsymbol{\gamma}}
     {\boldsymbol{\gamma}^T\Sigma_2'\boldsymbol{\gamma}}\leq x$ holds for arbitrary $\boldsymbol{\alpha},b$ when  $\boldsymbol{\gamma}=[\boldsymbol{\alpha},b]$.
     If it holds, then the Chernoff information of $N$ nodes trees is no more than $g(x)$, which is the result of $\boldsymbol{\gamma}^*=[\boldsymbol{\alpha}^*,0]$. So $\boldsymbol{\gamma}^*$ is the optimal observation and its $N$-st component is zero.

     From the structure of $\Sigma$, we can get this equation
      $\frac{\boldsymbol{\gamma}^T\Sigma_1'\boldsymbol{\gamma}}
     {\boldsymbol{\gamma}^T\Sigma_2'\boldsymbol{\gamma}}
     =\frac{\boldsymbol{\alpha}^T\Sigma_1\boldsymbol{\alpha}+b^2+2wb\boldsymbol{\alpha}^T\boldsymbol{\beta}_1
     }{\boldsymbol{\alpha}^T\Sigma_2\boldsymbol{\alpha}+b^2+2wb\boldsymbol{\alpha}^T\boldsymbol{\beta}_2
     }$.

     We consider two situations:

     1)$\alpha_j\neq 0$, we define $\boldsymbol{\theta}=\boldsymbol{\alpha}$ but replace $\theta_j$ with $0$. So

     \begin{align}
     \frac{1}{x}\leq\frac{\boldsymbol{\alpha}^T\Sigma_1\boldsymbol{\alpha}}
     {\boldsymbol{\alpha}^T\Sigma_2\boldsymbol{\alpha}}\leq x\\
     \frac{1}{x}\leq\frac{\boldsymbol{\theta}^T\Sigma_1\boldsymbol{\theta}}{\boldsymbol{\theta}^T\Sigma_2\boldsymbol{\theta}}
     =\frac{\boldsymbol{\alpha}^T\Sigma_1\boldsymbol{\alpha}+\alpha_j^2-2\alpha_j\boldsymbol{\alpha}^T\boldsymbol{\beta}_1
     }{\boldsymbol{\alpha}^T\Sigma_2\boldsymbol{\alpha}+\alpha_j^2-2\alpha_j\boldsymbol{\alpha}^T\boldsymbol{\beta}_2
     }\leq x
     \end{align}
     We know
     $
     \boldsymbol{\alpha}^T\Sigma_1\boldsymbol{\alpha}+b^2+2wb\boldsymbol{\alpha}^T\boldsymbol{\beta}_1=
     \frac{-wb}{\alpha_j}(\boldsymbol{\alpha}^T\Sigma_1\boldsymbol{\alpha}+\alpha_j^2-2\alpha_j\boldsymbol{\alpha}^T\boldsymbol{\beta}_1)
     +(1+\frac{wb}{\alpha_j})(\boldsymbol{\alpha}^T\Sigma_1\boldsymbol{\alpha})+(b^2+wb\alpha_j)
     $ and
      $
     \boldsymbol{\alpha}^T\Sigma_2\boldsymbol{\alpha}+b^2+2wb\boldsymbol{\alpha}^T\boldsymbol{\beta}_2=
     \frac{-wb}{\alpha_j}(\boldsymbol{\alpha}^T\Sigma_2\boldsymbol{\alpha}+\alpha_j^2-2\alpha_j\boldsymbol{\alpha}^T\boldsymbol{\beta}_2)
     +(1+\frac{wb}{\alpha_j})(\boldsymbol{\alpha}^T\Sigma_2\boldsymbol{\alpha})+(b^2+wb\alpha_j)
     $.
     So we can use proposition \ref{fact} and get
     \begin{align}
     \frac{1}{x}\leq\frac{\boldsymbol{\gamma}^T\Sigma_1'\boldsymbol{\gamma}}
     {\boldsymbol{\gamma}^T\Sigma_2'\boldsymbol{\gamma}}
     =\frac{\boldsymbol{\alpha}^T\Sigma_1\boldsymbol{\alpha}+b^2+2wb\boldsymbol{\alpha}^T\boldsymbol{\beta}_1
     }{\boldsymbol{\alpha}^T\Sigma_2\boldsymbol{\alpha}+b^2+2wb\boldsymbol{\alpha}^T\boldsymbol{\beta}_2
     }\leq x
     \end{align}

     2)$\alpha_j=0$, we define $\boldsymbol{\theta}=\boldsymbol{\alpha}$ but replace $\theta_j$ with $1$. So

     \begin{align}
      \frac{1}{x}\leq\frac{\boldsymbol{\alpha}^T\Sigma_1\boldsymbol{\alpha}}
     {\boldsymbol{\alpha}^T\Sigma_2\boldsymbol{\alpha}}\leq x
     \\
     \frac{1}{x}\leq\frac{\boldsymbol{\theta}^T\Sigma_1\boldsymbol{\theta}}{\boldsymbol{\theta}^T\Sigma_2\boldsymbol{\theta}}
     =\frac{\boldsymbol{\alpha}^T\Sigma_1\boldsymbol{\alpha}+1+2\boldsymbol{\alpha}^T\boldsymbol{\beta}_1
     }{\boldsymbol{\alpha}^T\Sigma_2\boldsymbol{\alpha}+1+2\boldsymbol{\alpha}^T\boldsymbol{\beta}_2
     }\leq x
     \end{align}

     So we can use proposition \ref{fact} and get
     \begin{align}
     \frac{1}{x}\leq\frac{\boldsymbol{\gamma}^T\Sigma_1'\boldsymbol{\gamma}}
     {\boldsymbol{\gamma}^T\Sigma_2'\boldsymbol{\gamma}}
     =\frac{\boldsymbol{\alpha}^T\Sigma_1\boldsymbol{\alpha}+b^2+2wb\boldsymbol{\alpha}^T\boldsymbol{\beta}_1
     }{\boldsymbol{\alpha}^T\Sigma_1\boldsymbol{\alpha}+b^2+2wb\boldsymbol{\alpha}^T\boldsymbol{\beta}_2
     }\leq x
     \end{align}

     So $
     \frac{1}{x}\leq\frac{\boldsymbol{\gamma}^T\Sigma_1'\boldsymbol{\gamma}}
     {\boldsymbol{\gamma}^T\Sigma_2'\boldsymbol{\gamma}}
     =\frac{\boldsymbol{\alpha}^T\Sigma_1\boldsymbol{\alpha}+b^2+2wb\boldsymbol{\alpha}^T\boldsymbol{\beta}_1
     }{\boldsymbol{\alpha}^T\Sigma_1\boldsymbol{\alpha}+b^2+2wb\boldsymbol{\alpha}^T\boldsymbol{\beta}_2
     }\leq x$ holds for all $\boldsymbol{\gamma}$, the proposition is proved.

\subsection{Node $i$ is $2$-degree node with the same neighbor}

     Without loss of generality, we make $i=N,w_{ip}=w_1,w_{iq}=w_2$.

    \begin{figure}
    \centering
    \includegraphics[width=3cm]{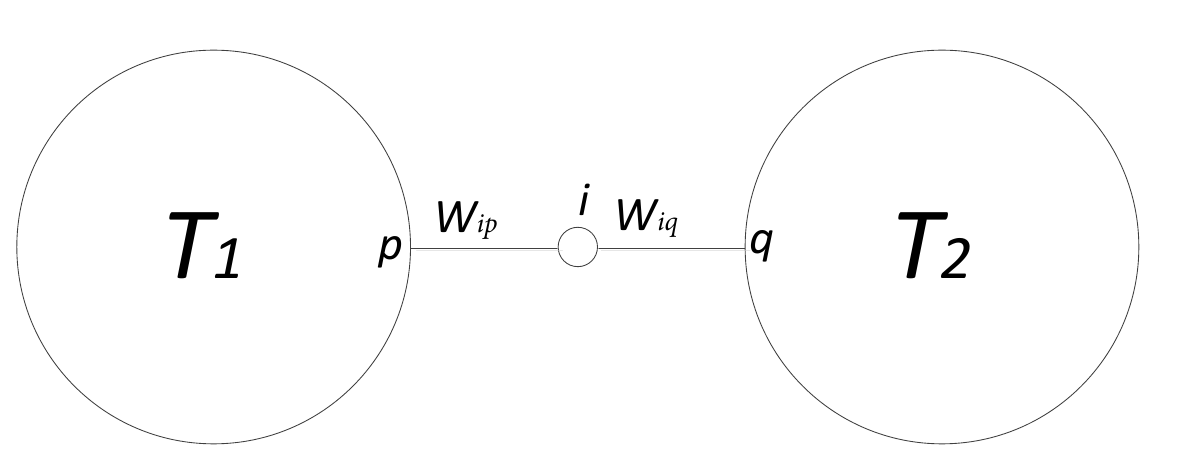}
    \includegraphics[width=3cm]{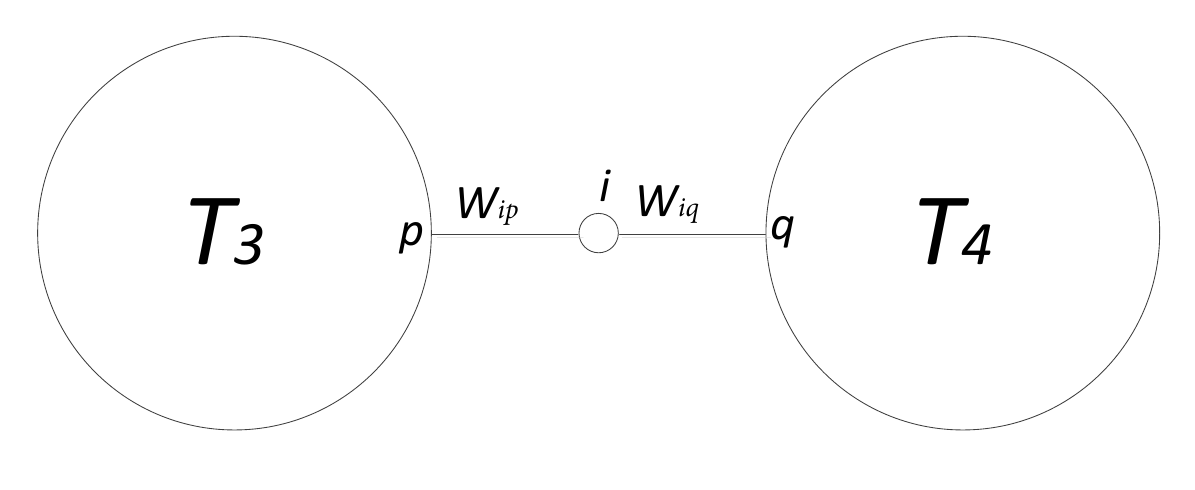}\\
    \caption{$2$-degree node $i$ in proposition ]ref{thm4}}\label{thm4-p2}
    \end{figure}

    The $N-1$ node trees without node $N$ have the normalized covariance matrix  $\Sigma_1$ and $\Sigma_2$.
    We get the $p$-st column of $\Sigma_1$ and set its $T_2$ nodes row(Fig.\ref{thm4-p2}) to be $0$ , namely $\boldsymbol{\beta}_{11}$. We get the $q$-st column of $\Sigma_1$ and set its $T_1$ nodes row to be $0$ , namely $\boldsymbol{\beta}_{12}$. So the $p$-st column of $\Sigma_1$ is $\boldsymbol{\beta}_{11}+w_1w_2\boldsymbol{\beta}_{12}$ and the q-st column of $\Sigma_1$ is $\boldsymbol{\beta}_{12}+w_1w_2\boldsymbol{\beta}_{11}$.  So as
    $\boldsymbol{\beta}_{21}$ and $\boldsymbol{\beta}_{22}$ are the same parameters with $\Sigma_2$.

     The complete trees can be treated as the $N-1$ node trees added with a new leaf $N$. So
     \begin{align}
     \Sigma_1'=\begin{bmatrix} \Sigma_1 &w_1\boldsymbol{\beta}_{11}+w_2\boldsymbol{\beta}_{12}\\w_1\boldsymbol{\beta}_{11}^T+w_2\boldsymbol{\beta}_{12}^T&1 \end{bmatrix}\\
     \Sigma_2'=\begin{bmatrix} \Sigma_2 &w_1\boldsymbol{\beta}_{21}+w_2\boldsymbol{\beta}_{22}\\w_1\boldsymbol{\beta}_{21}^T+w_2\boldsymbol{\beta}_{22}^T&1 \end{bmatrix}
     \end{align}

     We can find $x\geq1$ and $\boldsymbol{\alpha}^*$ so that $\frac{1}{x}\leq\frac{\boldsymbol{\alpha}^T\Sigma_1\boldsymbol{\alpha}}
     {\boldsymbol{\alpha}^T\Sigma_2\boldsymbol{\alpha}}\leq x$ holds for arbitrary $\boldsymbol{\alpha}$ and $\frac{\boldsymbol{\alpha}^{*T}\Sigma_1\boldsymbol{\alpha}^*}
     {\boldsymbol{\alpha}^{*T}\Sigma_2\boldsymbol{\alpha}^*}=x \:\text{or}\: \frac{1}{x}$. As (\ref{1dimCI}) shown,  $\boldsymbol{\alpha}^*$ is the optimal observation of the $N-1$-node trees and the optimal Chernoff information is $g(x)$.

     We want to prove that $\frac{1}{x}\leq\frac{\boldsymbol{\gamma}^T\Sigma_1'\boldsymbol{\gamma}}
     {\boldsymbol{\gamma}^T\Sigma_2'\boldsymbol{\gamma}}\leq x$ holds for arbitrary $\boldsymbol{\alpha},b$ when  $\boldsymbol{\gamma}=[\boldsymbol{\alpha},b]$.
     If it holds, then the Chernoff information of $N$ nodes trees is no more than $g(x)$, which is the result of $\boldsymbol{\gamma}^*=[\boldsymbol{\alpha}^*,0]$. So $\boldsymbol{\gamma}^*$ is the optimal observation and its $N$-st component is zero.

  From the structure of $\Sigma$, we can get this equation
    $\frac{\boldsymbol{\gamma}^T\Sigma_1'\boldsymbol{\gamma}}
     {\boldsymbol{\gamma}^T\Sigma_2'\boldsymbol{\gamma}}
     =\frac{\boldsymbol{\alpha}^T\Sigma_1\boldsymbol{\alpha}
     +b^2+2w_1b\boldsymbol{\alpha}^T\boldsymbol{\beta}_{11}+
     2w_2b\boldsymbol{\alpha}^T\boldsymbol{\beta}_{12}
     }{\boldsymbol{\alpha}^T\Sigma_2\boldsymbol{\alpha}
     +b^2+2w_1b\boldsymbol{\alpha}^T\boldsymbol{\beta}_{21}+
     2w_2b\boldsymbol{\alpha}^T\boldsymbol{\beta}_{22}
     }$.

     We consider three situations:

    1) $\alpha_p\neq 0,\alpha_q\neq 0$, we define $\boldsymbol{\theta}=\boldsymbol{\alpha}$ but replace $\theta_p$ with $0$, $\boldsymbol{\upsilon}=\boldsymbol{\alpha}$ but replace $\upsilon_q$ with $0$.

     \begin{align}
     \frac{1}{x}\leq\frac{\boldsymbol{\alpha}^T\Sigma_1\boldsymbol{\alpha}}
     {\boldsymbol{\alpha}^T\Sigma_2\boldsymbol{\alpha}}\leq x
     \\
     \frac{1}{x}\leq\frac{\boldsymbol{\theta}^T\Sigma_1\boldsymbol{\theta}}
     {\boldsymbol{\theta}^T\Sigma_2\boldsymbol{\theta}}
     =\frac{\boldsymbol{\alpha}^T\Sigma_1\boldsymbol{\alpha}+\alpha_p^2-2\alpha_p\boldsymbol{\alpha}^T\boldsymbol{\beta}_{11}
     -2w_1w_2\alpha_p\boldsymbol{\alpha}^T\boldsymbol{\beta}_{12}
     }{\boldsymbol{\alpha}^T\Sigma_2\boldsymbol{\alpha}+\alpha_p^2-2\alpha_p\boldsymbol{\alpha}^T\boldsymbol{\beta}_{21}
     -2w_1w_2\alpha_p\boldsymbol{\alpha}^T\boldsymbol{\beta}_{22}
     }\leq x
     \\
     \frac{1}{x}\leq\frac{\boldsymbol{\upsilon}^T\Sigma_1\boldsymbol{\upsilon}}
     {\boldsymbol{\upsilon}^T\Sigma_2\boldsymbol{\upsilon}}
     =\frac{\boldsymbol{\alpha}^T\Sigma_2\boldsymbol{\alpha}+\alpha_q^2-2w_1w_2\alpha_q\boldsymbol{\alpha}^T\boldsymbol{\beta}_{11}
     -2\alpha_q\boldsymbol{\alpha}^T\boldsymbol{\beta}_{12}
     }{\boldsymbol{\alpha}^T\Sigma_2\boldsymbol{\alpha}+\alpha_q^2-2w_1w_2\alpha_q\boldsymbol{\alpha}^T\boldsymbol{\beta}_{21}
     -2\alpha_q\boldsymbol{\alpha}^T\boldsymbol{\beta}_{22}
     }\leq x
     \end{align}

     So we can use proposition \ref{fact} and get
     \begin{align}
     \frac{1}{x}\leq\frac{\boldsymbol{\gamma}^T\Sigma_1'\boldsymbol{\gamma}}
     {\boldsymbol{\gamma}^T\Sigma_2'\boldsymbol{\gamma}}
     =\frac{\boldsymbol{\alpha}^T\Sigma_1\boldsymbol{\alpha}
     +b^2+2w_1b\boldsymbol{\alpha}^T\boldsymbol{\beta}_{11}+
     2w_2b\boldsymbol{\alpha}^T\boldsymbol{\beta}_{12}
     }{\boldsymbol{\alpha}^T\Sigma_2\boldsymbol{\alpha}
     +b^2+2w_1b\boldsymbol{\alpha}^T\boldsymbol{\beta}_{21}+
     2w_2b\boldsymbol{\alpha}^T\boldsymbol{\beta}_{22}
     }\leq x
     \end{align}

    2)$\alpha_p= 0,\alpha_q=0$,we define $\boldsymbol{\theta}=\boldsymbol{\alpha}$ but replace $\theta_p$ with $1$, $\boldsymbol{\upsilon}=\boldsymbol{\alpha}$ but replace $\upsilon_q$ with $1$.

    \begin{align}
     \frac{1}{x}\leq\frac{\boldsymbol{\alpha}^T\Sigma_1\boldsymbol{\alpha}}
     {\boldsymbol{\alpha}^T\Sigma_2\boldsymbol{\alpha}}\leq x
     \\
     \frac{1}{x}\leq\frac{\boldsymbol{\theta}^T\Sigma_1\boldsymbol{\theta}}
     {\boldsymbol{\theta}^T\Sigma_2\boldsymbol{\theta}}
     =\frac{\boldsymbol{\alpha}^T\Sigma_1\boldsymbol{\alpha}+1+2\boldsymbol{\alpha}^T\boldsymbol{\beta}_{11}
     +2w_1w_2\boldsymbol{\alpha}^T\boldsymbol{\beta}_{12}
     }{\boldsymbol{\alpha}^T\Sigma_2\boldsymbol{\alpha}+1+2\boldsymbol{\alpha}^T\boldsymbol{\beta}_{21}
     +2w_1w_2\boldsymbol{\alpha}^T\boldsymbol{\beta}_{22}
     }\leq x
     \\
     \frac{1}{x}\leq\frac{\boldsymbol{\upsilon}^T\Sigma_1\boldsymbol{\upsilon}}
     {\boldsymbol{\upsilon}^T\Sigma_2\boldsymbol{\upsilon}}
     =\frac{\boldsymbol{\alpha}^T\Sigma_2\boldsymbol{\alpha}+1+2w_1w_2\boldsymbol{\alpha}^T\boldsymbol{\beta}_{11}
     +2\boldsymbol{\alpha}^T\boldsymbol{\beta}_{12}
     }{\boldsymbol{\alpha}^T\Sigma_2\boldsymbol{\alpha}+1+2w_1w_2\boldsymbol{\alpha}^T\boldsymbol{\beta}_{21}
     +2\boldsymbol{\alpha}^T\boldsymbol{\beta}_{22}
     }\leq x
     \end{align}

     So we can use proposition \ref{fact} and get
     \begin{align}
     \frac{1}{x}\leq\frac{\boldsymbol{\gamma}^T\Sigma_1'\boldsymbol{\gamma}}
     {\boldsymbol{\gamma}^T\Sigma_2'\boldsymbol{\gamma}}
     =\frac{\boldsymbol{\alpha}^T\Sigma_1\boldsymbol{\alpha}
     +b^2+2w_1b\boldsymbol{\alpha}^T\boldsymbol{\beta}_{11}+
     2w_2b\boldsymbol{\alpha}^T\boldsymbol{\beta}_{12}
     }{\boldsymbol{\alpha}^T\Sigma_2\boldsymbol{\alpha}
     +b^2+2w_1b\boldsymbol{\alpha}^T\boldsymbol{\beta}_{21}+
     2w_2b\boldsymbol{\alpha}^T\boldsymbol{\beta}_{22}
     }\leq x
     \end{align}

     3)$\alpha_p= 0,\alpha_q\neq0$, we define $\boldsymbol{\theta}=\boldsymbol{\alpha}$ but replace $\theta_p$ with $1$, $\boldsymbol{\upsilon}=\boldsymbol{\alpha}$ but replace $\upsilon_q$ with $0$.($\alpha_p\neq 0,\alpha_q=0$ is the same)

     \begin{align}
      \frac{1}{x}\leq\frac{\boldsymbol{\alpha}^T\Sigma_1\boldsymbol{\alpha}}
     {\boldsymbol{\alpha}^T\Sigma_2\boldsymbol{\alpha}}\leq x
     \\
      \frac{1}{x}\leq\frac{\boldsymbol{\theta}^T\Sigma_1\boldsymbol{\theta}}
     {\boldsymbol{\theta}^T\Sigma_2\boldsymbol{\theta}}
     =\frac{\boldsymbol{\alpha}^T\Sigma_1\boldsymbol{\alpha}+1+2\boldsymbol{\alpha}^T\boldsymbol{\beta}_{11}
     +2w_1w_2\boldsymbol{\alpha}^T\boldsymbol{\beta}_{12}
     }{\boldsymbol{\alpha}^T\Sigma_2\boldsymbol{\alpha}+1+2\boldsymbol{\alpha}^T\boldsymbol{\beta}_{21}
     +2w_1w_2\boldsymbol{\alpha}^T\boldsymbol{\beta}_{22}
     }\leq x
     \\
     \frac{1}{x}\leq\frac{\boldsymbol{\upsilon}^T\Sigma_1\boldsymbol{\upsilon}}
     {\boldsymbol{\upsilon}^T\Sigma_2\boldsymbol{\upsilon}}
     =\frac{\boldsymbol{\alpha}^T\Sigma_2\boldsymbol{\alpha}+\alpha_q^2-2w_1w_2\alpha_q\boldsymbol{\alpha}^T\boldsymbol{\beta}_{11}
     -2\alpha_q\boldsymbol{\alpha}^T\boldsymbol{\beta}_{12}
     }{\boldsymbol{\alpha}^T\Sigma_2\boldsymbol{\alpha}+\alpha_q^2-2w_1w_2\alpha_q\boldsymbol{\alpha}^T\boldsymbol{\beta}_{21}
     -2\alpha_q\boldsymbol{\alpha}^T\boldsymbol{\beta}_{22}
     }\leq x
     \end{align}

     So we can use proposition \ref{fact} and get
     \begin{align}
     \frac{1}{x}\leq\frac{\boldsymbol{\gamma}^T\Sigma_1'\boldsymbol{\gamma}}
     {\boldsymbol{\gamma}^T\Sigma_2'\boldsymbol{\gamma}}
     =\frac{\boldsymbol{\alpha}^T\Sigma_1\boldsymbol{\alpha}
     +b^2+2w_1b\boldsymbol{\alpha}^T\boldsymbol{\beta}_{11}+
     2w_2b\boldsymbol{\alpha}^T\boldsymbol{\beta}_{12}
     }{\boldsymbol{\alpha}^T\Sigma_2\boldsymbol{\alpha}
     +b^2+2w_1b\boldsymbol{\alpha}^T\boldsymbol{\beta}_{21}+
     2w_2b\boldsymbol{\alpha}^T\boldsymbol{\beta}_{22}
     }\leq x
     \end{align}

     So $
     \frac{1}{x}\leq\frac{\boldsymbol{\gamma}^T\Sigma_1'\boldsymbol{\gamma}}
     {\boldsymbol{\gamma}^T\Sigma_2'\boldsymbol{\gamma}}
     =\frac{\boldsymbol{\alpha}^T\Sigma_1\boldsymbol{\alpha}
     +b^2+2w_1b\boldsymbol{\alpha}^T\boldsymbol{\beta}_{11}+
     2w_2b\boldsymbol{\alpha}^T\boldsymbol{\beta}_{12}
     }{\boldsymbol{\alpha}^T\Sigma_2\boldsymbol{\alpha}
     +b^2+2w_1b\boldsymbol{\alpha}^T\boldsymbol{\beta}_{21}+
     2w_2b\boldsymbol{\alpha}^T\boldsymbol{\beta}_{22}
     }\leq x$ holds for all $\boldsymbol{\gamma}$, the proposition is proved.

\section{Proof of Proposition~\ref{>1}}

    To prove this proposition, we construct a $p$-dim observation matrix $\mathbf{P}_{p\times m}=[\mathbf{Q}^*;\mathbf{K}_{(p-q)\times m}]$ whose output is $\mathbf{Y}_q$.

    When we use the observation $\mathbf{P}$, $\mathbf{Y}_q$ has $q$ values which are the same with $\mathbf{Y}_p^*$ and another $p-q$ different values.

    Using proposition \ref{combine-continous}, we know  \begin{align}CI(\mathbf{Y}_p^{(1)}||\mathbf{Y}_p^{(2)})\geq
    CI(\mathbf{Y}_q^{(1)*}||\mathbf{Y}_q^{(2)*}) \end{align}
    The definition of optimal matrix tells us \begin{align}CI(\mathbf{Y}_p^{(1)*}||\mathbf{Y}_p^{(2)*})\geq CI(\mathbf{Y}_p^{(1)}||\mathbf{Y}_p^{(2)})\end{align}
    So
    \begin{align}CI(\mathbf{Y}_p^{(1)*}||\mathbf{Y}_p^{(2)*})\geq CI(\mathbf{Y}_q^{(1)*}||\mathbf{Y}_q^{(2)*})\end{align}

\section{Calculation of Equation~\ref{calculate1}}

    The optimal $\boldsymbol{\alpha}^*$ is the one to maximize or minimize $\frac{\boldsymbol{\alpha}^T\Sigma_1\boldsymbol{\alpha}}
    {\boldsymbol{\alpha}^T\Sigma_2\boldsymbol{\alpha}}$.

     The Chernoff information of mapping $\boldsymbol{\alpha}$ and $k\boldsymbol{\alpha}$ is the same. When $\alpha_3=0$, the Chernoff information equals to $1$. So $\boldsymbol{\alpha}=[a_1,a_2,0]$ can't be the optimal observation. Therefore we only need consider $\boldsymbol{\alpha}=[a_1,a_2,1]$. So
    \begin{align}
    \boldsymbol{\alpha}^T\Sigma_1\boldsymbol{\alpha}
    =\boldsymbol{\alpha}^T\boldsymbol{\alpha}+2w_1a_1a_2+2w_2(w_1a_1+a_2)
    \\
    \boldsymbol{\alpha}^T\Sigma_2\boldsymbol{\alpha}
    =\boldsymbol{\alpha}^T\boldsymbol{\alpha}+2w_1a_1a_2+2w_2(a_1+w_1a_2)
    \end{align}

    We calculate the derivative of ratio first:
    \begin{align}
    \frac{\partial\frac{\boldsymbol{\alpha}^T\Sigma_1\boldsymbol{\alpha}}
    {\boldsymbol{\alpha}^T\Sigma_2\boldsymbol{\alpha}}}{\partial a_1}
    =&\frac{1}
    {{\big(\boldsymbol{\alpha}^T\Sigma_2\boldsymbol{\alpha}\big)}^2}
    \big\{2w_2(w_1-1)\big(2(w_1+1)w_2a_2
\nonumber\\&-a_1^2+a_2^2+1+2a_1a_2+2w_1a_2^2\big)\big\}
    \\
    \frac{\partial\frac{\boldsymbol{\alpha}^T\Sigma_1\boldsymbol{\alpha}}
    {\boldsymbol{\alpha}^T\Sigma_2\boldsymbol{\alpha}}}{\partial a_2}
    =&\frac{1}
    {{\big(\boldsymbol{\alpha}^T\Sigma_2\boldsymbol{\alpha}\big)}^2}
    \big\{2w_2(1-w_1)\big(2(w_1+1)w_2a_1
\nonumber\\&+a_1^2-a_2^2+1+2a_1a_2+2w_1a_1^2\big)\big\}
    \end{align}

    When $a_1,a_2\rightarrow\pm\infty$, $\frac{\boldsymbol{\alpha}^T\Sigma_1\boldsymbol{\alpha}}
    {\boldsymbol{\alpha}^T\Sigma_2\boldsymbol{\alpha}}=1$. So the optimal must be one of the stationary points, not at infinity.
    Stationary points means $\frac{\partial\frac{\boldsymbol{\alpha}^T\Sigma_1\boldsymbol{\alpha}}
    {\boldsymbol{\alpha}^T\Sigma_2\boldsymbol{\alpha}}}{\partial\boldsymbol{\alpha}}
    =0$. in other words,
    \begin{align}
    2(w_1+1)w_2a_2-a_1^2+a_2^2+1+2a_1a_2+2w_1a_2^2=&0\label{17}\\
    2(w_1+1)w_2a_1+a_1^2-a_2^2+1+2a_1a_2+2w_1a_1^2=&0\label{16}
    \end{align}
    Here we have deleted the points $w_2=0,a_3=0,w_1=1,a_1=a_2$ which lead to $\frac{\boldsymbol{\alpha}^T\Sigma_1\boldsymbol{\alpha}}
    {\boldsymbol{\alpha}^T\Sigma_2\boldsymbol{\alpha}}=1$ and can't be the optimization.

    Minus (\ref{17}) from (\ref{16}), we get
    \begin{align}
    2(w_1+1)(a_1-a_2)(w_2+a_1+a_2)=0
    \end{align}

    We ignore the point $a_1=a_2$, which stands for $\frac{\boldsymbol{\alpha}^T\Sigma_1\boldsymbol{\alpha}}
    {\boldsymbol{\alpha}^T\Sigma_2\boldsymbol{\alpha}}=1$.

    1) When $w_1=-1$, we calculate the equation set (\ref{17})(\ref{16}) and get the result $a_1-a_2=\pm 1$. The optimal Chernoff information is $g(\frac{1+|w_2|}{1-|w_2|})$ on the points $k[a_1,a_1\pm1,1]$.

    2) When $a_1=-{(w_2+a_2)}$, (\ref{17})(\ref{16}) becomes
    \begin{align}
    2(1-w_1)(w_2a_2+&a_2^2)-1+w_2^2=0\\
    a_1a_2=&\frac{1-w_2^2}{2w_1-2}
    \end{align}

    So $a_1,a_2$ satisfy $a_1+a_2=-w_2$ and $a_1a_2=\frac{1-w_2^2}{2w_1-2}$. Define $s_1,s_2$ is two solutions of equation $s^2+w_2s+\frac{1-w_2^2}{2w_1-2}=0$. So
    \begin{align}
    s_1=&-\frac{1}{2}(w_2+\sqrt{w_2^2-2\frac{1-w_2^2}{w_1-1}})\\
    s_2=&
    -\frac{1}{2}(w_2-\sqrt{w_2^2-2\frac{1-w_2^2}{w_1-1}})
    \end{align}
    The stationary points are $\boldsymbol{\alpha}_1=k[s_1,s_2,1],\boldsymbol{\alpha}_2=k[s_2,s_1,1]$ and the ratio $\frac{\boldsymbol{\alpha}_1^T\Sigma_1\boldsymbol{\alpha}_1}
    {\boldsymbol{\alpha}_1^T\Sigma_2\boldsymbol{\alpha}_1}$ are $C$ and $\frac{1}{C}$ where
    \begin{align}
    C=
    \frac{2-w_1w_2^2
    -w_2^2+(1-w_1)w_2\sqrt{w_2^2-2\frac{1-w_2^2}{w_1-1}}}
    {2-w_1w_2^2
    -w_2^2-(1-w_1)w_2\sqrt{w_2^2-2\frac{1-w_2^2}{w_1-1}}}
    \end{align}

    The function has only two stationary point, and $g(C)=g(\frac{1}{C})>1$. So the optimal points must be this two points. When $w_1=-1$, $s_1-s_2=-1$. So we can combine case 1) into case 2). So $\boldsymbol{\alpha}_1=k[s_1,s_2,1],\boldsymbol{\alpha}_2=k[s_2,s_1,1]$ are the optimal points all the time.

    In summary, when we can read only one value at a time, then the Chernoff information is:
    \begin{align}
    CI_1(\Sigma_1||\Sigma_2)=g(C)
    \end{align}
    where
    \begin{align}
    g(x)=&\frac{1}{2}\{\log\frac{x-1}{e\log x}+\frac{\log x}{x-1}\}\\
    C=&
    \frac{2-w_1w_2^2
    -w_2^2+(1-w_1)w_2\sqrt{w_2^2-2\frac{1-w_2^2}{w_1-1}}}
    {2-w_1w_2^2
    -w_2^2-(1-w_1)w_2\sqrt{w_2^2-2\frac{1-w_2^2}{w_1-1}}}
    \end{align}
    
    Here $C$ is equal to $\lambda_{max}$, and $s_1=-\frac{1}{2}\left(w_2+\sqrt{\beta}\right)$, $
    s_2=
    -\frac{1}{2}\left(w_2-\sqrt{\beta}\right)$.

\section{Calculation of Equation~\ref{calculate2}}

    We can find the exponential family between the two distribution $N(0,\Sigma_\lambda)$ where $\Sigma_\lambda^{-1}=\Sigma_1^{-1}\lambda+\Sigma_2^{-1}(1-\lambda)$.
    In order to calculate the Chernoff information, we have to find the equilibrium point $\lambda^*$ who satisfies $D(\Sigma_{\lambda^*}||\Sigma_2)=D(\Sigma_{\lambda^*}||\Sigma_1)$. The expression of the KL-distance is as shown in (\ref{D}). Simplify the equation and it becomes $tr(\Sigma_1^{-1}\Sigma_\lambda)=tr(\Sigma_2^{-1}\Sigma_\lambda)$. It's a simple equation and the result is $\lambda^*=\frac{1}{2}$. At last, we substitute it into $CI(\Sigma_1||\Sigma_2)=D(\Sigma_{\lambda^*}||\Sigma_2)$ and get the Chernoff information:
    \begin{align}
    CI_2(\Sigma_1||\Sigma_2)=\frac{1}{2}\log(1+\frac{1}{2}\frac{w_2^2}{1-w_2^2}(1-w_1))
    \end{align}
    Use $\lambda_{max}$ to place the parameters $(w_1,w_2)$, it becomes \ref{CI2}.

\section{Proof of Proposition~\ref{<2}}

    For equation \ref{g} and \ref{CI2}, we want to prove that $CI_1 \geq \frac{1}{2}G_2$ when $\lambda_{max}\geq1$. We can change the problem's form.
    
    Consider two functions $G_1(x)=\frac{\ln{x}}{x-1}-\ln{\frac{\ln{x}}{x-1}}-1$ and $G_2(x)=\ln{\frac{x+1}{2\sqrt{x}}}$. We only need prove $G_1(x)\geq G_2(x)$ for $x\in[1, \infty)$.

    \begin{align}
    G_1^{(1)}(x)=&\frac{(\ln{x}-x+1)(1-1/x-\ln{x})}{\ln{x}{(x-1)}^2}\\
    G_2^{(1)}(x)=&\frac{x-1}{2x(1+x)}\\
    F(x)=&2\ln{x}{(x-1)}^2x(x+1)\left(G_1^{(1)}(x)-G_2^{(1)}(x)\right)\nonumber\\
    =&2x(1+x)(\ln{x}-x+1)(1-1/x-\ln{x})-\ln{x}{(x-1)}^3\nonumber\\
    =&\ln{x}\left(x^3+5x^2-5x-1\right)-\ln^2{x}\left(2x+2x^2\right)-2x^3+2x^2+2x-2\\
    F^{(1)}(x)=&\ln{x}\left(3x^2+6x-9\right)-\ln^2{x}\left(4x+2\right)-5x^2+9x-3-1/x\\
    F^{(2)}(x)=&\ln{x}\left(6x-2-4/x\right)-4\ln^2{x}-7x+15-9/x+1/x^2\\
    F^{(3)}(x)=&\ln{x}\left(6-8/x+4/x^2\right)-1+5/x^2-2/x-2/x^3\\
    F^{(4)}(x)=&8\ln{x}\left(1-1/x\right)/x^2+6(1-1/x)(1-1/x^2)/x
    \end{align}

    $F^{(4)}(x)\geq0$ and $F^{(3)}(1)=0$, so $F^{(3)}(x)\geq0$.

    $F^{(3)}(x)\geq0$ and $F^{(2)}(1)=0$, so $F^{(2)}(x)\geq0$.

    $F^{(2)}(x)\geq0$ and $F^{(1)}(1)=0$, so $F^{(1)}(x)\geq0$.

    $F^{(1)}(x)\geq0$ and $F(1)=0$, so $F(x)\geq0$.

    $F(x)\geq0$, so $G_1^{(1)}(x)-G_2^{(1)}(x)\geq0$.

    $G_1^{(1)}(x)-G_2^{(1)}(x)\geq0$ and $G_1(x)-G_2(x)\overset{x\rightarrow1}{\rightarrow}0$, so $G_1(x)-G_2(x)\geq0$.

\end{document}